\documentclass[twocolumn,reprint,amsmath,amssymb,footinbib,superscriptaddress,floatfix,aps,longbibliography,pra]{revtex4-2}
\pdfoutput=1
\usepackage{hyperref}
\usepackage[utf8]{inputenc}
\usepackage[sort&compress]{natbib}
\usepackage[dvipsnames]{xcolor}
\usepackage{graphicx}
\usepackage{epsfig}
\usepackage{epstopdf}
\usepackage{amsfonts}
\usepackage{amstext,amsmath,amssymb}
\usepackage{svrsymbols}
\usepackage[normalem]{ulem}

\usepackage{rotating}
\usepackage{dcolumn}
\usepackage{array,multirow}
\usepackage[english]{babel}
\usepackage[export]{adjustbox}
\usepackage{braket}
\usepackage{xr}
\usepackage{ifpdf}
\usepackage{tikz}
\usetikzlibrary{arrows}

\hypersetup{urlcolor=violet, citecolor=violet, linkcolor=violet, colorlinks=true}
\newcommand{\be}{\begin{equation}}
\newcommand{\ee}{\end{equation}}
\newcommand{\bea}{\begin{eqnarray}}
\newcommand{\eea}{\end{eqnarray}}

\graphicspath{{draft/figs/}}

\begin{document}

\title{Spectrum analysis with parametrically modulated transmon qubits}

\author{Nir Gavrielov}
\affiliation{Racah Institute of Physics, Hebrew University of Jerusalem, 91904 Jerusalem, Israel}
\email{nir.gavrielov@mail.huji.ac.il}

\author{Santiago Oviedo-Casado}
\affiliation{\'Area de F\'isica Aplicada, Universidad Polit\'ecnica de Cartagena, Cartagena E-30202, Spain}
\affiliation{Escuela Superior de Ingeniería y Tecnología, Universidad Internacional de La Rioja, 26006, Logroño, La Rioja, Spain}
\email{santiago.oviedo@unir.net}

\author{Alex Retzker}
\affiliation{Racah Institute of Physics, Hebrew University of Jerusalem, 91904 Jerusalem, Israel}
\affiliation{AWS Center for Quantum Computing, Pasadena, California 91125, USA}
\email{retzker@phys.huji.ac.il}

\begin{abstract}
Exploring the noise spectrum impacting a qubit and extending its coherence duration are fundamental components of quantum technologies. In this study, we introduce parametric spectroscopy, a method that merges parametric modulation of a qubit's energy gap with dynamical decoupling sequences. The parametric modulation provides high sensitivity to extensive regions of the noise spectrum, while dynamical decoupling reduces the effect of driving noise. Our theoretical study shows that parametric spectroscopy enables access to the difficult high-frequency domain of the flux spectrum in transmons.
\end{abstract}

\maketitle

\section{Introduction}

Preserving quantum states long enough to realize quantum operations is a challenging task. Noise, which causes internal system relaxation and the reduction of coherence, is one of the major hurdles that aspiring quantum technologies face \cite{Preskill2018}, 
and has sparkled a myriad of noise mitigation techniques based on sophisticated forms of dynamical decoupling (DD) \cite{hahn1950spin,carr1954effects,meiboom1958modified,viola1998dynamical,Viola1999,dephasing_Das_Sarma,Souza2011,Suter2016}. The basic working principle of DD is to create customized frequency filters that reduce the harmful impact of uncontrolled degrees of freedom \cite{kofman2001universal,kofman2004unified,Biercuk2011}, resulting in longer coherence times. Owing to this frequency selectivity, the same protocols can, in principle, be used for noise spectroscopy \cite{Alvarez2011,Bylander2011,yuge2011measurement,young2012qubits}.

Superconducting qubits (SQs) are one of the main platforms to enact quantum technologies, in which high-fidelity gate operations and long coherence times are routinely achieved. 
A precise and detailed knowledge of the noise spectrum and its impact on the system is crucial, and for that reason a whole body of work has tried to unveil the different noise sources affecting superconducting qubits \cite{Krantz2019,Yan2013,ithier2005decoherence, sank2012flux, yan2012spectroscopy,Bylander2011}. However, transmons find it difficult to access the high-frequency region of the noise power spectrum and can generally only achieve frequencies up to a few hundred MHz, thereby leaving large portions of the spectrum inadequately explored. It is possible to reach the high frequency range by tuning the energy gap\cite{Bylander2011,astafiev2004quantum}; however, such a technique suffers from low signal-to-noise ratio (SNR). Another option is to utilize spin locking on the flux qubit, which has significant nonlinearity, to attain Rabi frequencies in the GHz range \cite{yoshihara2014flux,Yan2013}.

Interestingly, relaxometry analysis of the high-frequency noise spectrum displays an Ohmic behavior around the SQ natural energy gap of 5-6 GHz \cite{Yan2016}, in stark contrast with the 1/$f$ spectrum observed at low frequencies \cite{paladino2014_1f}. Methods for reliably accessing the unknown regions of the SQ noise spectrum are imperatively needed.

Among SQ families, the tunable transmon features a flexible energy gap that can be parametrically modulated to resonate with another SQ using an oscillating radio-frequency flux \cite{Bertet2006,PhysRevApplied.10.034050,Niskanen2006}, thereby favoring gate creation with great flexibility and state-of-the-art fidelity \cite{Bertet2006,Niskanen2006, Niskanen2007, Beaudoin2012, Royer2017,Roy2017, McKay2016, Reagor2018, Hong2020}. The transmon possesses static and dynamical sweet spots, specific energy gaps at which the transmon becomes relatively insensitive to flux noise, attaining long coherence times \cite{Krantz2019,rigetti_ac_sweet}. Here, we aim to utilize the fact that during frequency modulation around the static sweet spots, tunable transmons behave as frequency filters, potentially serving as accurate noise spectrometers while maintaining long coherence.

In this article, we suggest integrating microwave pulsed dynamical decoupling sequences within parametric modulation, enabling spectrum sampling at any desired modulation frequency. This approach facilitates high-resolution noise spectroscopy across a broad spectral range, from a few MHz to GHz. We begin by introducing the fundamental concept of \emph{parametric spectroscopy}, demonstrating its operating principle in a tunable transmon. Subsequently, we theoretically and numerically exhibit unmatched spectral resolution at the high-frequency end of the spectrum, a region that has previously resisted most probing efforts. Moreover, we show a substantial improvement in coherence time along with minimized leakage to higher energy levels. The versatility of the protocols we investigate implies that parametric spectroscopy can be utilized with any qubit possessing a tunable energy gap. This opens new opportunities for noise spectroscopy across various setups and provides a novel tool for achieving extended coherence times, an essential resource for quantum technologies.

\begin{figure*}
    \centering
    \includegraphics[width=\linewidth]{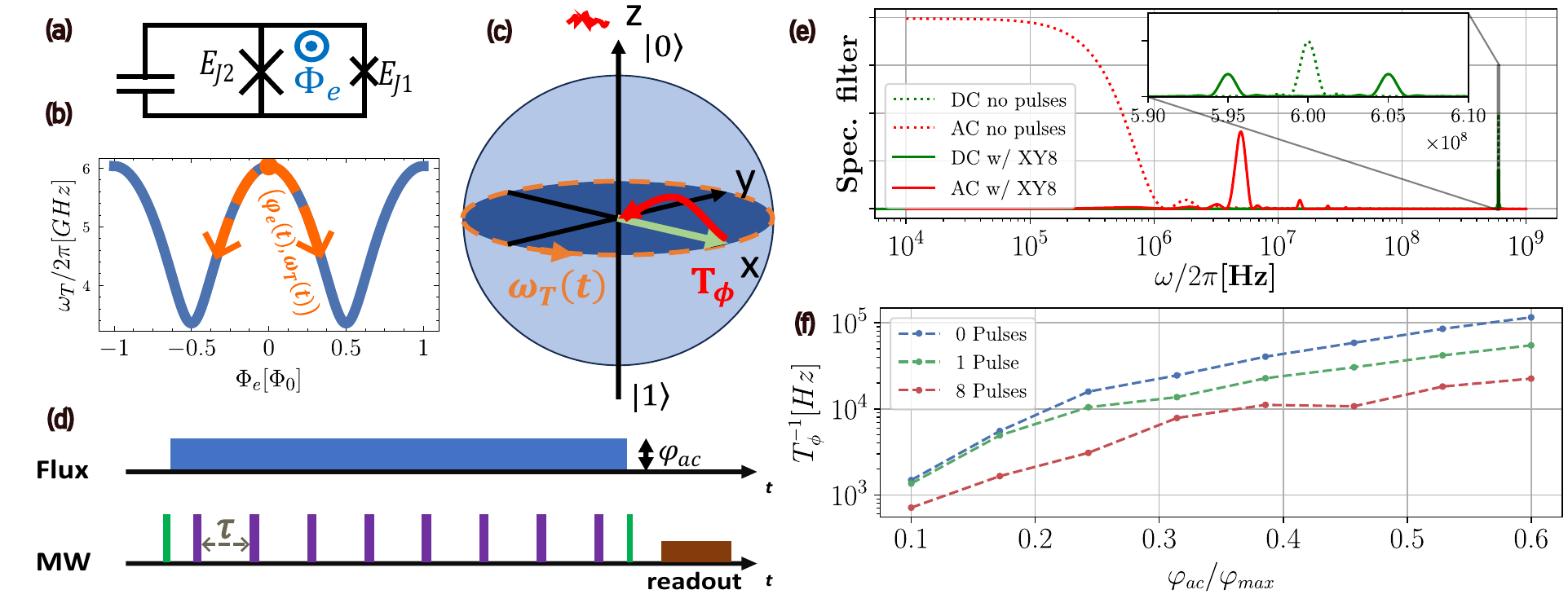}
    \caption{(a) Schematic tunable transmon circuit. (b) Qubit energy curve, with an illustration of parametric modulation (orange) of the transmon energy gap. (c) Bloch sphere depiction of dephasing during parametric modulation. The Bloch vector (green) of a superposition state rotates around the polarization axis at an angular frequency that matches the energy gap $\omega_{T}$. Longitudinally coupled noise (red) causes loss of phase information. Tuning of the energy gap through $\omega_m$ permits matching $\omega_T$ to specific noise frequencies that are detected. (d) Sketch of a parametric spectroscopy sequence with a pulse train on both the flux and the microwave channels. (e) Spectral filters for the additive (DC, green) and multiplicative (AC, red) noise sources during parametric spectroscopy. Solid lines show the filters when an additional $XY$8 sequence is applied, shifted from the original parametric modulation spectrum (dashed). (f) Dephasing rate $T_\phi^{-1}$ of the modulated qubit ($\frac{\omega_m}{2\pi}\approx500MHz$) from numerical simulations, where $\varphi_{max}=\pi/2$ marks the first minimum of $\omega_T\left(\varphi_e\right)$. Adding concurrent dynamical decoupling pulses filters out unwanted noisy frequencies, with the immediate consequence of decreasing the dephasing rate, as shown by the application of a single $\pi$ rotation (Hahn echo), as compared to parametric modulation without DD pulses (dashed blue). Applying an $XY$8 DD sequence provides up to a fivefold decrease in the dephasing rate in a point-by-point comparison with bare parametric modulation. Such a decrease amounts to a fivefold increase in coherence time when compared to state-of-the-art parametrically modulated transmons. \cite{Hong2020, Valery2022}}
    \label{fig:Figure1}
\end{figure*}

\section{Parametric spectroscopy} 

We begin by describing the method operating in a transmon device, considered for now as a two-level system with an energy gap $E_{01} = \hbar\omega_T$.  The qubit frequency $\omega_T$ is periodically controlled by an external flux $\Phi_{e}$ threading the superconducting quantum interference device (SQUID) loop shown in Fig.~\ref{fig:Figure1}(a) \cite{koch_transmon}. Parametric modulation of the transmon stands for applying a sinusoidal drive,
\begin{equation} \label{eq:flux_drive}
\varphi_{e}\left(t\right)=\varphi_{dc}+\delta\varphi_{dc}\left(t\right)+\left[\varphi_{ac}+\delta\varphi_{ac}\left(t\right)\right]\sin\left(\omega_{m}t\right),
\end{equation}
  of modulation frequency $\omega_m$, through the flux channel, with $\varphi_{e}\equiv\pi\Phi_{e}/\Phi_{0}$, normalized to the flux quantum unit $\Phi_0$. The qubit response around a sweet spot--shown in Fig.~\ref{fig:Figure1}(b)--is, approximately, quadratic: 
\begin{equation}        \omega_{T}\left(\varphi_{e}\right)\simeq\omega_{T}\left(0\right)+b_{0}\varphi_{e}^{2}, \qquad b_{0}\equiv\frac{1}{2}\frac{\partial^{2}\omega_{T}\left(\varphi_{e}\right)}{\partial\varphi_{e}^{2}}, 
\end{equation}
in which $b_0$, the curvature around the peak, depends on the capacitive ($E_{C}$) and junction ($E_{J1}, E_{J2}$) energies of the circuit. The net result is that the qubit becomes maximally sensitive to additive environmental noise frequencies that match the modulated qubit energy gap, henceforth DC noise $\delta\varphi_{dc}$, while being desensitized to the rest of the environment spectrum. Thus, for typical transmon parameters, the sensitivity of parametric modulation to a pure tone signal is on par--$2\varphi_{ac}/3$--with that of an optimal spin-locking sequence (see Appendix~\ref{sec:Sensitivity}).  On the other hand, it means that the qubit is additionally subjected to multiplicative AC noise $\delta\varphi_{ac}$ introduced by the flux modulation amplitude.

Interaction with noise causes qubit frequency fluctuations which, when the system is prepared in an initial superposition state, result in dephasing, whose characteristic time $T_\phi$ is defined through the decoherence function $W\left(t\right)\equiv\left|\left\langle\exp\left( -i\intop_{0}^{t}\delta\omega_{T}\left(t^{\prime}\right)dt^{\prime}\right)\right\rangle \right|$ as $W\left(T_\phi\right)=1/\text{e}$ \cite{dephasing_Das_Sarma}. Throughout this work, we assume Gaussian noise and ignore possible noise relaxation effects, focusing on the $T_1\gg T_\phi$ regime \cite{paladino2014_1f}. The coupling of the noise terms to an oscillating signal leads to a frequency shift in the spectral domain, as shown in Fig.~\ref{fig:Figure1}(e). As a result, the dephasing of the qubit depends on the spectrum of the environmental noise at the modulation frequency,
\begin{multline}
\label{eq:decoherence}
    -\ln W\left(t\right)\simeq b_{0}^{2}\varphi_{ac}^{2}t\left(S_{dc}\left(\omega_{m}\right)+\frac{1}{4}S_{ac}\left(2\omega_{m}\right)\right) + \\
    \frac{b_{0}^{2}\varphi_{ac}^{2}t^{2}}{2}\intop_{-\infty}^{\infty}\frac{d\omega}{2\pi}{\text{sinc}}^{2}\left(\frac{\omega t}{2}\right)S_{ac}\left(\omega\right),
\end{multline}
with $S\left(\omega\right)$ the power spectral density at a frequency $\omega$. A detailed derivation of the decoherence function can be found in Appendix~\ref{appendix: decohernce-integrals}.


Experimental evidence suggests a strong decoherence contribution from electronic (multiplicative) noise \cite{Reagor2018}, which typically has a 1/$f$ spectrum with lower (upper) $\omega_{ir}  \left(\omega_{uv}\right)$ cutoff, which we assume from now on. Then, the zero-frequency multiplicative noise [the last term in Eq.~(\ref{eq:decoherence})] overshadows the target signal $S_{dc}\left(\omega_{m}\right)$, as depicted in Fig.~\ref{fig:Figure1}(e), causing low SNR for frequency detection through parametric modulation, limiting its scope to low frequencies that require small modulation amplitudes. We overcome the difficulty by concurrently applying a microwave pulsed DD sequence that reduces the impact of the multiplicative process while keeping the sensitivity to the additive target signal intact. Considering a series of equidistant impulsive $\pi$ rotations separated by a delay time $\tau$, as sketched in Fig.~\ref{fig:Figure1}(d), causes the additive and multiplicative spectral filters due to parametric modulation to split onto two windows, each shifted by $\pm\,\omega^{\prime}=\pi /\tau$ from its original central frequency [see Fig.~\ref{fig:Figure1}(e)] \cite{dephasing_Das_Sarma, Bylander2011}. Typical DD sequences shift the frequency by a few MHz, pushing the multiplicative noise away from the origin, but only slightly affecting the parametric filter, which is at least tens of MHz.  Consequently, the filter becomes increasingly precise, permitting larger modulation amplitudes that allow reaching the high-frequency spectrum and achieving, at the same time, a significant reduction of dephasing, as we demonstrate in Fig.~\ref{fig:Figure1}(f) (see also Appendix~\ref{appendix: decohernce-integrals}). There, we consider both a Hahn echo (single $\pi$ rotation) and the paradigmatic $XY$8 DD sequence, which minimizes possible pulse errors by alternating the $\pi$ rotations around the different axes of the Bloch sphere, thereby partially compensating the error of a given pulse by the other pulses in the sequence \cite{Souza2011}. The addition of this $XY$8 DD sequence to parametric modulation offers a fivefold increase in the dephasing time $T_\phi$ compared to bare parametric modulation, owing to the frequency selectivity introduced by the DD sequence, which significantly reduces the impact of the noisy environment. The extended dephasing time results in further improvement in the sensitivity, 
\begin{equation}
     \label{eq: sensitivity}
S_{dc_{min}}^{p}\left(\omega_{m}\right)\approx\frac{\text{e}}{2Cb_0^2\varphi_{ac}^{2}\sqrt{T_{\phi}}}
\end{equation}
where $C$ is the overall readout efficiency parameter. A better trade-off between the strength of the noise signal (through $\varphi_{ac}$) and the coherence time can be achieved to obtain state-of-the-art sensitivities when targeting signals in a noisy environment \cite{Degen2017}. In what follows, we demonstrate analytically and numerically enhanced resolution of frequencies at the high-energy end of the spectrum of a tunable transmon device.

\section{Resolution gain}

We now turn to analyze the spectral resolution properties of \emph{parametric spectroscopy}, benchmarked by the minimal distance that two spectral peaks must have in order to be distinguished. We illustrate the analytical calculation for the case of frequency estimation, which is equivalent to estimating the distance between the peaks if the target is the average frequency \cite{Gefen2019}. Thus, we aim to determine the central frequency $\omega_c$ of a single peaked Gaussian spectral line through relaxometry with a parametric spectroscopy sequence on a tunable transmon. The decay rate measured due to the additive signal, $\Gamma \equiv b_0^2\varphi_{ac}^2S_{dc}(\omega_m)$, is directly related to the noise spectrum through Eq.~(\ref{eq:decoherence}), which includes the target spectral line and the background pink noise. Then, any frequency estimation is limited by $\sigma_\Gamma$, the uncertainty in the estimation of $\Gamma$.

The experimental procedure consists of a series of binary measurements on the transmon, where the probability of observing the qubit in its $\ket{0}$ state depends on the decoherence function and reads $p\left(\Gamma,t\right)=e^{-\Gamma t-\left(\alpha t\right)^{2}}$, with $\alpha=b_{0}\varphi_{ac} A_{ac}\sqrt{\ln\left(e\omega_{uv}/\omega_{ir}\right)}$ due to the multiplicative noise source. To quantify the experimental precision in estimating $\Gamma$, we use the root mean squared error $\sigma_\Gamma$, which, according to the Cram{\'e}r-Rao bound, is lower bound by the inverse of the Fisher information for said parameter \cite{Cramer1946,Rao1992}. Considering measurements of optimal duration $T_\phi$, and assuming that the multiplicative, AC decay rate $\alpha$ is known and constant for different modulation frequencies, as predicted by the theoretical analysis in Eq.~(\ref{eq:decoherence}), we get that  
\begin{equation} 
    I\left(\Gamma\right)=\left\langle \left(\frac{\partial\ln{\mathcal{L}}}{\partial\Gamma}\right)^{2}\right\rangle=\frac{T_\phi^2}{e^{\Gamma T_\phi+\left(\alpha T_\phi\right)^{2}}-1}, 
\end{equation}
with $\mathcal{L}$ the likelihood function for the probabilistic model for $\Gamma$. To calculate the precision in estimating $\Gamma$, we consider an experiment of total duration $T_{\text{meas}}$, consisting of $N$ measurements each of duration $T_\phi$ and with a (small) overhead time $t_m$ for probe initialization and readout \cite{sensing_review}. Then, using the additivity of the Fisher information, we get that
\be
\sigma_{\Gamma}=\sqrt{\frac{T_\phi+t_{m}}{T_{\text{meas}}}}\sqrt{\frac{1}{I\left(\Gamma\right)}} \approx \sqrt{\frac{1}{T_{\text{meas}}T_\phi}}.
\ee

The goal is to estimate $\omega_c$. Assuming enough independent identical repetitions, the decay rate estimator is normally distributed. Then, the likelihood that a given decay rate $x$ was measured for a given spectral feature with a peak frequency $\omega_c$ is equal to
\begin{equation}
    \mathcal{L}=\frac{1}{\sigma_{\Gamma}\sqrt{2\pi}}\exp\left[-\frac{1}{2\sigma_{\Gamma}^{2}}\left(x-\frac{A_{\omega}}{\sigma_{\omega}\sqrt{2\pi}}e^{-\frac{\left(\omega-\omega_{c}\right)^{2}}{2\sigma_{\omega}^{2}}}\right)^{2}\right],
\end{equation}
where $\omega$ is the sampled frequency and $\sigma_{\omega} \sim T_\phi^{-1}$
\cite{Barry2020}. In addition, $A_{\omega}$ is the amplitude of the fluctuations of the energy gap due to the additive noise at the spectral peak. Then, the uncertainty of the frequency estimation is
$\Delta\left(\omega_{c}\right)\geq 1/\sqrt{I(\omega_c)}$, with 
\begin{equation}
     I\left({\omega_c}\right)=\left\langle \left(\frac{\partial\ln\mathcal{L}}{\partial\omega_{c}}\right)^{2}\right\rangle=\frac{A_{\omega}^{2}e^{-\frac{\left(\delta\omega\right)^{2}}{\sigma_{\omega}^{2}}}}{2\pi\sigma_{\Gamma}^{2}\sigma_{\omega}^{6}}\left(\delta\omega\right)^{2},
\end{equation}
where $\delta\omega\equiv\omega-\omega_{c}$. Choosing the appropriate detuning $\delta_\omega = \sigma_\omega$ that maximizes the Fisher information about the frequency, and repeating the procedure for $N_\omega$ sequences each of $T_{\text{meas}}$ duration, yields a total $I_{\text{tot}}^{\omega_c}=\sum_{k=1}^{N_{\omega}}I_{k}\mid\delta\omega_{k}= \sigma_\omega$, such that the uncertainty of the estimation is
\begin{equation} \label{eq:FI_one_freq} \Delta\left(\omega_{c}\right) \geq 
    \left(\frac{2\pi\text{e}}{A_{\omega}^{2}N_\omega T_{\text{meas}} T_\phi^5 } \right)^{\frac{1}{2}}.
\end{equation}

\begin{figure}[t!]
    \centering
    \includegraphics[width=\columnwidth]{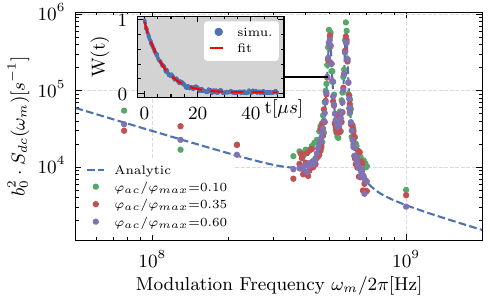}
    \caption{Simulation of a full spectroscopy. The noise that is generated includes a pink spectrum \cite{pink_noise_nasa} $A_{\Phi}/w$  with amplitude $A_{\Phi}\approx\left(60\mu\Phi_{0}\right)^{2}$ and two shifted AR1 processes with coherence time $T_{AR1}=25ns$ and standard deviation $\sigma_{\Phi_{e}}\approx\left(247\mu\Phi_{0}\right)$. The centers of the AR1 spectra are located at 500 and 580MHz, matching a distance of twice the full width at half maximum (FWHM) between the peaks. The decoherence function is estimated by averaging $2^{11}$ trials of numerical integration of the accumulated phase with the modulation signal, an ideal $XY$8 sequence, and the described noise. Finally, the decay rate for the specific parameters is extracted using a Gaussian and exponential fit (e.g., in the inset for $\varphi_{ac}=0.6\varphi_{max};\omega_{m}/2\pi\approx489MHz$) and scaled to the flux noise density using Eq~(\ref{eq:decoherence}). We scanned 86 modulation frequencies in the $10^7-10^9$ Hz range and three modulation amplitudes. The sampled points are compared to the analytical result (dashed line) showing excellent spectrum reconstruction.}
    \label{fig:PSD}
\end{figure}

The procedure for frequency resolution is formally equivalent: assuming knowledge of the mean frequency, the spectral distance $\epsilon\equiv\omega^{\left(1\right)}_{c}-\omega^{\left(2\right)}_{c}$ becomes the target frequency parameter \cite{Gefen2019,Rotem2019}, derivation details can be found in Appendix~\ref{resolution}. 
There are two possible scenarios: When $\epsilon > T_\phi^{-1}$, the uncertainty is that from estimation, given by Eq.~(\ref{eq:FI_one_freq}), and resolving the frequencies is, in principle, a matter of performing enough measurements. Conversely, when $\epsilon\ll T_\phi^{-1}$, choosing the optimal detuning, which in this case corresponds to a modulation frequency $\omega_m = \left(\omega^{\left(1\right)}_{c}+\omega^{\left(2\right)}_{c}\right)/2$, yields a resolution uncertainty that approaches
\begin{equation}    \Delta\left(\epsilon\right) \geq \left(\frac{2\pi}{A_{\omega}^{2}N_\omega T_{\text{meas}} T_\phi^7 \epsilon^2} \right)^{\frac{1}{2}} = \frac{\Delta(\omega_c) }{T_\phi \epsilon}\frac{1}{\sqrt{\text{e}}} ,\label{eq:resolution}
\end{equation}
showing that resolution is limited by the finite coherence time of the probe. 
The advantage of parametric spectroscopy comes from its ability to achieve long coherence times for high frequencies. Therefore, uncertainties as small as 100 kHz even for GHz frequencies are within reach.
Further improvement of the resolution limit imposed by Eq.~(\ref{eq:resolution}) should be possible by considering sequential measurement schemes such as Qdyne \cite{Schmitt2017,Degen2017,Glenn2018}, which optimize spectral resolution in noisy environments and dispel the problem of the limited coherence time of the probe \cite{Gefen2019,Staudenmaier2023}. A parametric Qdyne experiment targeting the mean frequency between two peaks would yield a signal oscillating at an $\epsilon$ frequency. Then, a sequence of evenly time-separated measurements reflects an oscillation at the peak separation frequency. For spectral features whose coherence time is longer than $T_\phi$, the single-frequency limit is thus recovered, such that $\Delta(\epsilon) \gtrsim \left(A_{\omega}^{2} T_\phi^4 T_{\text{meas}}^2 \right)^{-1/2}$.

Figure~\ref{fig:PSD} shows a numerical analysis of frequency resolution, in the high-frequency region of the spectrum, with parametric spectroscopy. We evaluate
$W(t)$ from Eq.~(\ref{eq:decoherence}) for various modulation frequencies $\omega_{m}$ and amplitudes $\varphi_{ac}$, including an $XY$8 DD sequence, and both additive and multiplicative noise sources modeled as a pink noise spectrum of amplitudes $A^{dc}_\Phi\approx\left(35\mu\Phi_{0}\right)^{2}/\omega$ and $A_{\Phi}^{ac}$ matching $\sqrt{\left\langle \delta\varphi_{ac}^{2}\right\rangle }/\varphi_{ac}=0.004\%$. Additionally, we include two Gaussian peaks with central frequencies $\omega_c^{\left(1\right)} = 500$ MHz and $\omega_c^{\left(2\right)} = 580$ MHz, both with a full width at half maximum (FWHM) of 40 MHz. 
Spectral reconstruction is performed by averaging, for each sampling point, $2^{11}$ realizations of the accumulated phase at the end of the sequence, and then fitting that average to a Gaussian and exponential decay form. The exponential decay rate estimator quantifies the additive source dephasing rate, showing excellent agreement with the analytical prediction (see Fig.~\ref{fig:PSD_simu_details} in Appendix~\ref{sec:simu_deets}). Further, we demonstrate resolution of the central frequencies of the Gaussian features by fitting the reconstructed spectrum around the peaks to the analytical model, leaving the central frequencies, widths, and amplitudes as free fitting parameters. The resolution procedure for $\{\omega_c^{\left(1\right)},\omega_c^{\left(2\right)}\}$ involves taking the average of 100 fits with a minimum required $R^2 > 0.8$, yielding $\omega^{\left(1\right)}_{c}/2\pi = 500MHz, \omega^{\left(2\right)}_{c}/2\pi = 570 MHz$ both with confidence bounds of $\Delta\left(\omega^{\left(i\right)}_{c}\right)/2\pi\approx14MHz$, well below the FWHM of the peaks. Note that applying more stringent conditions to the fitting would permit approaching the fundamental limit imposed by the Fisher information. 

The results displayed in Fig.~\ref{fig:PSD} assume an infinite $T_1$ relaxation time. However, for high frequencies, it might be the case that $2 T_1 < T_\phi$. Then, $T_1$ becomes the limiting factor for resolution, replacing $T_\phi$ in Eq.~(\ref{eq:resolution}). In general, post selection can be applied to mitigate the effect of a short $T_1$ by choosing to measure dephasing only for those sample measurements that did not relax. This however requires a special erasure qubit construction \cite{Kubica_2023,Levine_2024}. The price to pay is an increase in the experiment time $T_{\text{meas}}$ necessary to achieve the same resolution, often prohibitively so. In practice, for large $2T_1 \gg T_\phi$ post selection is not necessary, as relaxation is minimal (see Fig.~\ref{fig:T1} in Appendix~\ref{sec:t1_effect}) and results are equivalent to those with infinite $T_1$ in Fig.~\ref{fig:PSD}. On the other end, a very short $2T_1\ll T_\phi$ causes severe degradation of the ability to estimate the peaks. Note that relaxation and dephasing are, in principle, unrelated processes, for which reason, even if for a short $T_1$ resolution is severely compromised, parametric spectroscopy still offers prolonged $T_\phi$ time.

\section{Leakage mitigation} 

The two-level system approximation for superconducting qubits is limited by the unavoidable presence of higher-energy levels. Finite time drives inevitably address these unwanted levels, causing population leakage from the computational (sensing) manifold and decreasing the fidelity of operations. Leakage represents a major drawback for microwave-based operations on transmons, where it limits the gate speed imposing a minimum width for the pulses and prevents the qubit sensor from being able to accurately probe the high-energy end of the spectrum. Here, we theoretically and numerically analyze leakage during parametric spectroscopy. Our starting point is the full cosine Hamiltonian of a tunable transmon \cite{koch_transmon,Krantz2019},
\be 
\tilde{H}=4E_{c}n^{2}-E_{J_{eff}}\left(\varphi_{e}\right)\cos\left(\varphi\right)-\dot{\varphi}_{0}\left(\varphi_{e}\right)n,
\label{eq:full_cosine}
\ee
with $E_{J,eff}\left(\varphi_{e}\right)=\left(E_{J1}+E_{J2}\right)\sqrt{\cos^{2}\left(\varphi_{e}\right)+d^{2}\sin^{2}\left(\varphi_{e}\right)}$ being the effective, tunable Josephson energy, and $n$ the charge operator. In the presence of a fast flux drive, the last term in Eq.~(\ref{eq:full_cosine}) is non-negligible and causes transitions to higher levels. Around the operational point $\varphi = 0$, the phase operator $\varphi$ resembles that of a bosonic field, and the full cosine Hamiltonian Eq.~(\ref{eq:full_cosine}) can be approximated by an anharmonic oscillator. The resulting Hamiltonian, including the dynamic flux drive, reads
\begin{equation} \label{eq:Kerr_Ham}
    H=\omega_{T}\left(\varphi_{e}\left(t\right)\right)a^{\dagger}a-\frac{\eta}{2}a^{\dagger2}a^{2}-\frac{\dot{\varphi}_{0}\left(\varphi_{e}\left(t\right)\right)i}{2\sqrt{\xi}}\left(a^{\dagger}-a\right),
\end{equation}
with $a^{\dagger} \left(a\right)$ the creation (annihilation) operator, $\xi\simeq\sqrt{2E_c/\left(E_{J1}+E_{J2}\right)}$ and $\varphi_0=\tan{\left[d\cdot tan^{-1}{\left(\varphi_{e}\right)}\right]}\simeq d \varphi_{e}$ under the quadratic approximation. $\eta > 0$ in Eq.~(\ref{eq:Kerr_Ham}) represents the anharmonicity of the transmon.

Transition to higher-energy levels is apparent in a frame rotating with the dynamic part of the qubit frequency, $\omega_{T}\left(\varphi_{e}\left(t\right)\right)-\bar{\omega}_{T}$, where $\bar{\omega}_{T}=\frac{1}{T}\intop_{0}^{T}\omega_{T}\left(t\right)dt$ is the average energy gap during a period $T=\left(2\pi\right)/\omega_{m}$. The Hamiltonian in this frame, during a period of flux modulation as described in Eq.~(\ref{eq:flux_drive}), reads
\begin{multline} \label{eq: I_Ham}
    H_{I}=\bar{\omega}_{T}a^{\dagger}a-\frac{\eta}{2}a^{\dagger2}a^{2}-\frac{id\varphi_{ac}\omega_{m}}{2\sqrt{\xi}}\times \\
    \sum_{n=-\infty}^{\infty}J_{n}\left(\frac{\tilde{\omega}_{T}}{2\omega_{m}}\right)\left[e^{i\left(2n+1\right)\omega_{m}t}a^{\dagger}-e^{-i\left(2n+1\right)\omega_{m}t}a\right]
\end{multline}
where $J_{n}$ is the first Bessel function of order $n$ and $\tilde{\omega}_{T}=b_{0}\cdot\varphi_{ac}^{2}/2$ is the oscillation amplitude in the qubit frequency. The Hamiltonian in Eq.~(\ref{eq: I_Ham}) describes an anharmonic oscillator of frequency $\bar{\omega}_{T}$ perturbed by harmonic interactions with magnitudes of the Bessel series. 
For relevant sequence parameters, the resonance is determined by $n_{res}=\left[\left(\left(\bar{\omega_{T}}-\eta\right)/\omega_{m}-1\right)/2\right]>\tilde{\omega}_{T}/\left(2\omega_{m}\right)$. Then, transitions are strongly suppressed by the properties of the Bessel function $J_{n}\left(z\right)\underset{n\gg z}{\sim}\left(z/2\right)^{n}/n!$ for positive $n$. Furthermore, the contribution from dispersive interaction from low-$n$ terms can be bounded; specifically, the Rabi-like amplitude for the $\left|1\right\rangle \rightarrow\left|2\right\rangle $ transition is lower than $g^2/\left(g^2+\delta^2\right)$ with detuning and coupling
\begin{equation} \label{eq:leak_rabi}
    \delta\equiv\bar{\omega}_{T}-\eta-\omega_{m}\qquad g=d\omega_{m}\varphi_{ac}\sqrt{2/\xi}
\end{equation}

\begin{figure}[t]
    \centering
    \includegraphics[width=\columnwidth]{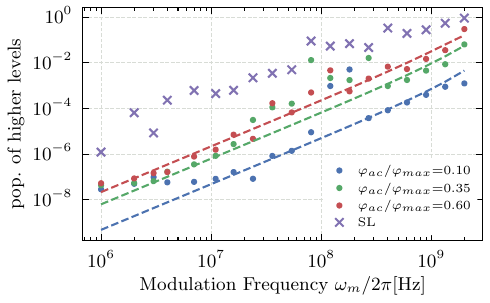}
    \caption{Parametric spectroscopy leakage analysis. Results of simulated (dots) 10$\mu s$ modulation and $XY$8 pulse sequence, applied to the full cosine tunable transmon Hamiltonian~(\ref{eq:full_cosine}), with $\omega_{max}/2\pi=6$ GHz, $\eta/2\pi=300$ MHz and $d\approx1/3$. Each dot represents the average maximal population outside the $\left\{ \left|0\right\rangle ,\left|1\right\rangle \right\}$ (computational or sensing) subspace during the last 10 time steps (0.1 ns) of the simulated dynamics (see Appendix~\ref{sec:simu_deets}), thus mitigating phase variability of the many active multi-
    level interactions. The dashed lines correspond to the analytical result for parametric modulation without DD from Eq.~(\ref{eq: I_Ham}) with the parameters from Eq.~(\ref{eq:leak_rabi}). 
    The violet $\times$ marks represent the numerical results for an analog spin-locking Hamiltonian (the horizontal axis being the Rabi frequency $\Omega$). Simulations were performed using QuTiP \cite{qutip}.}
    \label{fig:leak_sim}
\end{figure}

Figure \ref{fig:leak_sim} numerically analyzes the leakage for parametric spectroscopy with an $XY$8 sequence on a transmon, simulated by a diagonalized full cosine Hamiltonian projected onto the 10 lowest eigenstates. Significantly lower leakage rates with respect to spin locking, here representing a generic microwave drive procedure, are observed. This is particularly noteworthy at high frequencies, where other methods display a close to one probability of finding the population at higher levels and where, even for large modulation amplitude, parametric spectroscopy leakage rates are less than 1 \% of those from conventional protocols. The analytical result without pulses in Eq.~(\ref{eq: I_Ham}) with Eq.~(\ref{eq:leak_rabi}) parameters describes well the full sequence leakage results when we apply DD sequence leakage mitigation through derivative-removal-by-adiabatic-gate \cite{drag1, drag2}, together with pulse shape optimization through maximization of the average fidelity of quantum operations \cite{NIELSEN2002249}. Note that the slight bump appearing around 100 MHz originates on a resonance due to the DD pulses and can be addressed via a more thorough pulse optimization, tailored to the specific system that is considered. The slight \textit{plateau} present at low frequencies is a numerical artifact of the full cosine simulation, i.e., the temporal propagation of a dynamical flux procedure with fixed flux numerically diagonalized $n,\varphi$ operators introduces additional leakage that is not expected to be observed in an experiment. These results justify the two-level approximation employed to demonstrate high resolution with parametric spectroscopy.

\section{Discussion} 

Full spectral reconstruction of the noise interacting with a quantum system is critical for the development of quantum technologies. Here we demonstrate how tunable transmons serve as precise quantum sensors of the high-frequency noise spectrum. Combining parametric modulation with dynamical decoupling allows the creation of custom narrow frequency filters capable of extending the frequency range of the tunable transmon spectrometer into the GHz region, with unrivaled frequency resolution. We show great adaptability across the spectrum, increased coherence survival, and a significant reduction in leakage to higher-energy levels. 

Any sensing protocol requires a trade-off between the amplitude of the control pulses, the target frequency, the desired resolution, and leakage minimization. Unlike alternative strategies that require large modulation amplitudes that shorten coherence time and degrade attainable resolution \cite{rigetti_ac_sweet, Hong2020, Huang2021, Valery2022}, parametric spectroscopy has the unique advantage of allowing greater flexibility with innately built noise resilience when operated at close vicinity to a sweet spot. Moreover, it benefits from noise mitigation strategies that can be directly applied to the DD sequence, permitting larger amplitudes with reduced leakage and, thereby, increasing the sensitivity at high frequencies without compromising the spectrometer. Further, the protocol that we describe can be used to address structures in the spectrum reflected through noise correlations following a similar procedure to Ref.~\cite{OliverCorrelations} (additional details can be found in Appendix~\ref{sec:correlation}), and can be generalized to any system holding tunable energy gaps and therefore has the potential to become a useful addition to the quantum sensing toolbox. 

\section*{Acknowledgements} The contents of this article have greatly benefited from fruitful discussions with Ido Zuk and Yotam Vaknin. S.O.-C. acknowledges support from the María Zambrano Fellowship. A.R. acknowledges the support of ERC grant QRES, project number 770929, Quantera grant MfQDS, the Israeli Science foundation (ISF), the Schwartzmann university chair and the Israeli Innovation Authority under the project "Quantum Computing Infrastructures".

\newpage
\appendix
\onecolumngrid

\section{Decoherence Integral} \label{appendix: decohernce-integrals}
In this appendix, we present the derivation of the decoherence function under parametric modulation and within the quadratic approximation, Eq.~(\ref{eq:decoherence}) in the main text. We begin with the Hamiltonian of a two-level system qubit subjected to longitudinal noise, $H=\left[\omega_{T}+\delta\omega_{T}\left(t\right)\right]\sigma_{z}/2$. The characteristic dephasing time $T_\phi$ for a superposition state on the qubit is calculated from the \emph{decoherence function}, defined as
\begin{equation}
\label{eq:deco-def}
    W\left(t\right)\equiv\left|\left\langle \exp\left(-i\intop_{0}^{t}\delta\omega_{T}\left(t^{\prime}\right)dt^{\prime}\right)\right\rangle \right|,
\end{equation}
such that $W\left(T_\phi\right) = 1/e $, i.e. $T_\phi$ corresponds with the FWHM of the power spectrum. In the case of parametric modulation, an oscillating flux signal  $\varphi_{e}\left(t\right)=\varphi_{dc}+\delta\varphi_{dc}\left(t\right)+\left[\varphi_{ac}+\delta\varphi_{ac}\left(t\right)\right]\sin\left(\omega_{m}t\right)$ threads the superconducting quantum interference device (SQUID) loop of a tunable flux qubit. Additive and multiplicative flux noise sources cause energy gap fluctuations. Then, the noise at the resonance frequency of the qubit can be obtained from
\begin{equation}
 \delta\omega_{T}\left(t\right)=\frac{\partial\omega_{T}}{\partial\varphi_{dc}}\bigg|_{\varphi_{dc}}\delta\varphi_{dc}+\frac{\partial\omega_{T}}{\partial\varphi_{ac}}\bigg|_{\varphi_{ac}}\delta\varphi_{ac}.
\end{equation}
The resonant energy response close to one of the flux protected sweet-spots of the device is, approximately, quadratic: 
\begin{equation}        \omega_{T}\left(\varphi_{e}\right)\simeq\omega_{T}\left(0\right)+b_{0}\varphi_{e}^{2}, \qquad b_{0}\equiv\frac{1}{2}\frac{\partial^{2}\omega_{T}\left(\varphi_{e}\right)}{\partial\varphi_{e}^{2}}, 
\end{equation}
which means that the fluctuations of the energy gap can be effectively approximated by
\begin{equation}
\delta\omega_{T}\left(t\right)\simeq\left[-2b_{0}\varphi_{ac}\sin\left(\omega_{m}t\right)\right]\delta\varphi_{dc}\left(t\right)
-\left\{b_{0}\varphi_{ac}\left[1-\cos\left(2\omega_{m}t\right)\right]\right\}\delta\varphi_{ac}\left(t\right),
\end{equation}
which we can now use to calculate the decoherence function defined in Eq.~(\ref{eq:deco-def}). Under the common Gaussian noise assumption $W\left(t\right)=\exp\left[-\gamma_\phi\left(t\right)\right]$, with 
\begin{align}
\begin{split}
\gamma_\phi\left(t\right) &=\frac{1}{2} \intop_{0}^{t}dt_{1}\intop_{0}^{t}dt_{2}\left\langle \delta\omega_T\left(t_1\right)\delta\omega_T\left(t_2\right)\right\rangle = \\&
2b_{0}^{2}\varphi_{ac}^{2}\intop_{0}^{t}dt_{1}\intop_{0}^{t}dt_{2}\left\langle \delta\varphi_{dc}\left(t_{1}\right)\delta\varphi_{dc}\left(t_{2}\right)\right\rangle\sin\left(\omega_{m}t_{1}\right)\sin\left(\omega_{m}t_{2}\right) \\&
     +\frac{b_{0}^{2}\varphi_{ac}^{2}}{2}\intop_{0}^{t}dt_{1}\intop_{0}^{t}dt_{2}\left\langle \delta\varphi_{ac}\left(t_{1}\right)\delta\varphi_{ac}\left(t_{2}\right)\right\rangle
     \left[1-\cos\left(2\omega_{m}t_{1}\right)-\cos\left(2\omega_{m}t_{2}\right)+\cos\left(2\omega_{m}t_{1}\right)\cos\left(2\omega_{m}t_{2}\right)\right].
     \label{eq:5}
     \end{split}
\end{align} 
To write Eq.~(\ref{eq:5}) we take the reasonable assumption that the additive noise source (originating in the environment) is uncorrelated with the multiplicative noise source (electronically originating), that is, $\left\langle \delta\varphi_{ac}\left(t_{1}\right)\delta\varphi_{dc}\left(t_{2}\right)\right\rangle=0$. 
Then, the additive part --the first integral in Eq.~(\ref{eq:5})-- is equal to
\begin{equation}  \label{eq:add_decoher}
    \begin{split}
    & \intop_{0}^{t}dt_{1}\intop_{0}^{t}dt_{2}\sin\left(\omega_{m}t_{1}\right)\sin\left(\omega_{m}t_{2}\right)\underset{C_{\varphi}\left(t_{1}-t_{2}\right)}{\underbrace{\left\langle \delta\varphi_{dc}\left(t_{1}\right)\delta\varphi_{dc}\left(t_{2}\right)\right\rangle}} \\ = & \intop_{0}^{t}dt_{1}\intop_{0}^{t}dt_{2}\intop_{-\infty}^{\infty}\frac{d\omega}{2\pi}S_{dc}\left(\omega\right)e^{i\omega\left(t_{1}- t_{2}\right)}\sin\left(\omega_{m}t_{1}\right)\sin\left(\omega_{m}t_{2}\right) \\ =
    & \frac{1}{2}\intop_{-\infty}^{\infty}d\omega\left[\frac{\sin^{2}\left(\left(\omega+\omega_{m}\right)\frac{t}{2}\right)}{\pi\left(\omega+\omega_{m}\right)^{2}}+\frac{\sin^{2}\left(\left(\omega-\omega_{m}\right)\frac{t}{2}\right)}{\pi\left(\omega-\omega_{m}\right)^{2}}\right]S_{dc}\left(\omega\right) + \frac{\cos\left(\omega_{m}t\right)}{\pi}\intop_{-\infty}^{\infty}d\omega S_{dc}\left(\omega\right)\frac{\sin\left(\frac{\omega+\omega_{m}}{2}t\right)\sin\left(\frac{\omega-\omega_{m}}{2}t\right)}{\left(\omega+\omega_{m}\right)\left(\omega-\omega_{m}\right)} 
    \\ \rightarrow & \frac{t}{4}\intop_{-\infty}^{\infty}S_{dc}\left(\omega\right)\left[\delta\left(\omega+\omega_{m}\right)+\delta\left(\omega-\omega_{m}\right)\right] + \frac{\cos\left(\omega_{m}t\right)}{\pi}\intop_{-\infty}^{\infty}S_{dc}\left(\omega\right)\delta\left(\omega+\omega_{m}\right)\delta\left(\omega-\omega_{m}\right) \\ \simeq & \frac{t}{2}S_{dc}\left(\omega_{m}\right),
    \end{split}
\end{equation}
where we also make the (safe) assumption that the correlation time of the noise is much shorter than $t$. Similar calculations lead to all the other decay terms present contributing to Eq.~(\ref{eq:decoherence}) in the main text. 

\emph{Adding dynamical decoupling pulses} --- Let us now consider the effect that adding dynamical decoupling pulses simultaneously with the flux drive has on the decoherence function.
When a series of $2n-1$ equally spaced impulsive $\pi$ rotations is added during the experiment, the terms inside the integral in Eq.~(\ref{eq:5}) change sign every time a pulse arrives. Then, Eq.~(\ref{eq:add_decoher}) transforms into 
\begin{align}
\begin{split}
    &\frac{1}{8\pi}\intop_{-\infty}^{\infty}d\omega S_{dc}\left(\omega\right) \times \\& 
    \left|\frac{1}{\left(\omega+\omega_{m}\right)}\sum_{k=0}^{2n-1}\left(-1\right)^{k}\left[e^{\frac{i\left(\omega+\omega_{m}\right)\left(k+1\right)}{2n}}-e^{\frac{i\left(\omega+\omega_{m}\right)k}{2n}}\right] - \frac{1}{\left(\omega-\omega_{m}\right)}\sum_{k=0}^{2n-1}\left(-1\right)^{k}\left[e^{\frac{i\left(\omega-\omega_{m}\right)\left(k+1\right)}{2n}}-e^{\frac{i\left(\omega-\omega_{m}\right)k}{2n}}\right]\right|^{2},
\end{split}
\end{align}
which can be simplified to yield
\begin{equation}
    \simeq\frac{1}{2\pi}\intop_{-\infty}^{\infty}d\omega S_{dc}\left(\omega\right)\left[\frac{\tan^{2}\left(\frac{\omega+\omega_{m}}{4n}t\right)\sin^{2}\left(\frac{\omega+\omega_{m}}{2}t\right)}{\left(\omega+\omega_{m}\right)^{2}}+\frac{\tan^{2}\left(\frac{\omega-\omega_{m}}{4n}t\right)\sin^{2}\left(\frac{\omega-\omega_{m}}{2}t\right)}{\left(\omega-\omega_{m}\right)^{2}}\right].
\end{equation}
The term in brackets describes a frequency filter composed of 4 peaks centered at $\omega=\pm\omega_{m}\pm2\pi\frac{2n-1}{2t}$.

\section{\label{sec:Sensitivity} Sensitivity}
In this appendix, we analyze the \textit{sensitivity} of parametric modulation and compare it with that of spin-locking. Assuming a coherent magnetic flux signal $\delta\varphi_{dc}\left(t\right)=A_{\varphi,dc}\cos\left(\omega_{0}t\right)$ that threads the SQUID loop in addition to the parametric modulation signal $\varphi_{e}\left(t\right)=\varphi_{ac}\cos\left(\omega_{m}t\right)$, we question how well can the amplitude $A_{\varphi,dc}$ be estimated \cite{sensing_review}. The phase accumulated by the probe during this protocol is given by 
\begin{equation}
    \exp\left(-i\intop_{0}^{t}\omega_{T}\left(t^{\prime}\right)dt^{\prime}\right) = e^{-i\omega_{max}t}\exp\left\{ -ib_{0}\intop_{0}^{t}dt^{\prime}\left[\varphi_{ac}\cos\left(\omega_{m}t^{\prime}\right)+A_{\varphi,dc}\cos\left(\omega_{0}t^{\prime}\right)\right]^{2}\right\},
\end{equation}
where the second equality is valid under the quadratic approximation $\omega_{T}\left(t\right)\simeq\omega_{T}\left(0\right)+b_{0}\varphi_{e}^{2}\left(t\right)$, denoting $\omega_{T}\left(0\right)\equiv\omega_{max}$. The maximum sensitivity is reached when we probe at the frequency of the target signal $\omega_m=\omega_0$, yielding 
\begin{equation}
    \approx e^{-i\omega_{max}t}\exp\left[-ib_{0}\frac{t}{2}\left(\varphi_{ac}+A_{\varphi,dc}\right)^{2}\right].
\end{equation}
After the known contributions $\omega_{max}, \varphi_{ac}$ are removed, the signal amplitude can then be estimated from the oscillations' frequency
\begin{equation}
    S_{p}\approx b_{0}\varphi_{ac}A_{\varphi,dc}.
\end{equation} 
As a benchmark, we compare the sensitivity of parametric spectroscopy to the sensitivity of spin-locking spectroscopy to an equivalent signal. The Hamiltonian of a tunable transmon qubit under a spin-locking drive reads
\begin{equation}
    H_{SL}=\left[\frac{\omega_{q}}{2}+A_{SL}\cos\left(\omega_0 t\right)\right]\sigma_{z}+\Omega\cos\left(\omega_{q}t\right)\sigma_{x},
\end{equation}
where $A_{SL}\cos\left(\omega_{0}t\right)$ are the fluctuations due to the flux signal, which become maximal on the slope at $\varphi_e=\pi/4$, where 
\begin{equation}
    A_{SL}\simeq\frac{\partial\omega_{T}}{\partial\varphi_{e}}\mid_{\pi/4}A_{\varphi,dc}=\left[2/\left(1+d^2\right)\right]^{3/4}b_{0}A_{\varphi,dc}\approx\frac{3}{2}b_{0}A_{\varphi,dc},
\end{equation}
for common asymmetry values $d\approx1/3$ \cite{Krantz2019}. In the rotating frame, when $\omega_0=\Omega$, and under the rotating-wave approximation $\omega_q\gg\Omega$ 
\begin{equation}
    \tilde{H}_{SL}=\frac{\Omega}{2}\sigma_{x}+A_{SL}\cos\left(\Omega t\right)\sigma_{z}.
\end{equation}
On the Bloch sphere, when we set the initial state as $\left|0_{x}\right\rangle$, oscillations around the Z axis are observed, with a frequency 
\begin{equation}
    A_{SL}\approx\frac{3}{2}b_{0}A_{\varphi,dc}.
\end{equation}
Therefore, both methods can achieve similar sensitivity, as 
\begin{equation}
    \frac{S_{p}}{S_{SL}}=\frac{\nu_{2}\left(0\right)\varphi_{ac}}{\nu_{1}\left(\frac{\pi}{4}\right)}\approx\frac{2}{3}\varphi_{ac} \lesssim 1,
\end{equation}
where $\nu_{k}\left(\varphi_{e}\right)=\frac{1}{k!}\frac{\partial^{k}\omega_{T}}{\partial\varphi_{e}^{k}}|_{\varphi_{e}}$ is k'th power series coefficient of $\omega_T\left(\varphi_e\right)$.

For noisy signals, we use the definition of sensitivity as the minimal magnitude of the signal that produces a state probability SNR$_p$ = 1 during $T_{meas}$=1s \cite{Degen2017}, SNR standing for signal-to-noise ratio. In this framework 
\be
SNR_p=\frac{\delta p_{obs}}{\sigma_{p}}=\delta p\left(t_{o}\right)e^{-\chi\left(t_{o}\right)}2C\sqrt{\frac{T_{meas}}{t_{o}+t_{m}}},
\ee
with $\delta p, \sigma_p$ the measured qubit phase and its uncertainty, $\chi\left(t\right)$ the qubit decoherence, $t_{o}$ the interrogation time, C a factor of readout fidelity and $t_{m}$ the initialization and readout time. In parametric spectroscopy the target noise signal is equal to $\varphi_{ac}^{2}b_{0}^{2}S_{dc}\left(\omega_{m}\right)t_o$ and the decoherence is derived from Eq.~(\ref{eq: sensitivity}) in the main text. Substituting these into the definition above, under the $t_o\approx T_\phi, t_m\ll t_o$ assumptions, yields a minimal signal of 
\be
S_{dc_{min}}^{p}\left(\omega_{m}\right)\approx\frac{e}{2Cb_0^2\varphi_{ac}^{2}\sqrt{T_{\phi}}}.
\ee
For an equivalent magnetic flux noise, the rotating frame relaxation in an spin-locking experiment will follow $\chi\left(t_o\right) = \nu_{1}^2\left(\pi/4\right)S_{dc}\left(\Omega\right)t_{o}$,
where $\Omega$ is the Rabi frequency and $\nu_1\left(\pi/4\right)$ is the flux derivative on the slope. Similarly, the minimal detectable signal for this procedure is equal to
\be
S_{dc_{min}}^{SL}\left(\Omega\right)\approx\frac{e}{2C\nu_{1}^2\left(\pi/4\right)\sqrt{T_{1\rho}}},
\ee
with $T_{1\rho}$ the relaxation time of the Spin Locked qubit. Consequently, the parametric procedure can show a gain in sensitivity up to 
\be
\frac{S_{dc_{min}}^{p}\left(\omega_{m}\right)}{S_{dc_{min}}^{SL}\left(\Omega\right)}\approx\frac{9}{4\varphi_{ac}^{2}}\sqrt{\frac{T_{1\rho}}{T_{\phi}}}.
\ee

\section{\label{sec:SNR} SNR analysis}

\begin{figure}[h]
            \includegraphics[width=\textwidth,height=\textheight,keepaspectratio]{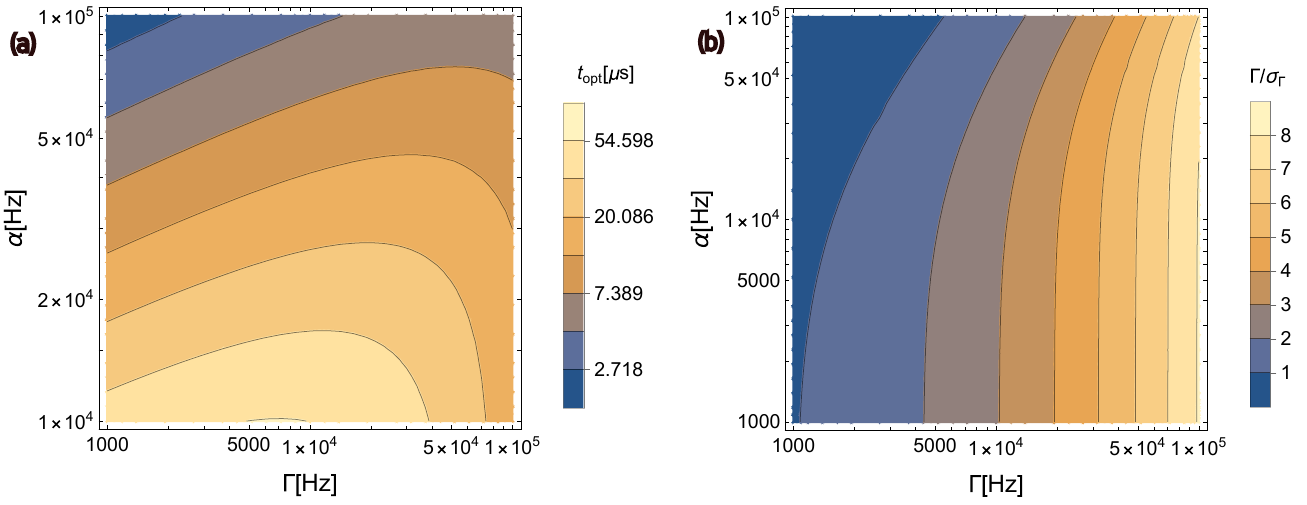}
    \caption{Precision Analysis. (a) Optimal measurement time, defined as the measurement time that maximizes Fisher information, Eq.~(\ref{eq:param_fisher}), for different values of the decay rate $\Gamma$. For values of $\Gamma>\alpha$, the optimal timing follows $\Gamma^{-1}$ as expected from single exponential decay [see Eq.~(\ref{eq:fish_exp})]. However, when $\alpha>\Gamma$ the multiplicative decay masks the signal and $t_{opt}$ follows primarily $\alpha^{-1}$. (b) Plot of the measurement SNR, given a total experiment time of $T_{meas}=1ms$ and trial overhead of $t_{m}=1\mu s$ \cite{Walter2017, Tuorila2017, Sunada2022}, showing that $SNR > 1$ is feasible.}
    \label{fig:fisher}
\end{figure}

To set a quantitative number for the standard error around a measured signal, we use standard information theory methods. Our simulated experimental procedure consists of a series of binary measurements performed on the qubit probe, with the probability of state measurement $\left|0\right\rangle$ set by the decoherence function. We can therefore model such a procedure as a sequence of Bernoulli trials, each with a time-dependent success probability 
\begin{equation}
    p\left(\Gamma,t\right)=e^{-\Gamma t-\left(\alpha t\right)^{2}},
\end{equation}
where $\Gamma=b_{0}^{2}\varphi_{ac}^{2}S_{dc}\left(\omega_{m}\right)$ is the additive noise target signal, and $\alpha=b_{0}\varphi_{ac} A_{ac}\sqrt{\ln\left(e\omega_{uv}/\omega_{ir}\right)}$ comes from the  multiplicative source, which has 1/$f$ spectrum and tends to mask the target signal. The log-likelihood function for this probabilistic model is 
\begin{equation}
    \ln{\mathcal{L}}\left(\Gamma\mid x\right)=x\log p+\left(1-x\right)\log\left(1-p\right),
\end{equation}
where the possible values for the random variable x are $\left\{0,1\right\}$. Since we are interested in estimating $\Gamma$, we calculate the Fisher information obtained about the parameter $\Gamma$ from a given signal trial, which is
\begin{equation} \label{eq:param_fisher}
    I\left(\Gamma,t\right)=\left\langle \left(\frac{\partial\log{\mathcal{L}}}{\partial\Gamma}\right)^{2}\right\rangle=\left(\frac{\partial p}{\partial\Gamma}\right)^{2}\frac{1}{p\left(1-p\right)}=t^{2}\frac{1}{e^{\Gamma t+\left(\alpha t\right)^{2}}-1},
\end{equation}
where we assume that the multiplicative, AC decay rate $\alpha$ is known during the experiment, as the theoretical analysis predicts a constant rate for different modulation frequencies. Since the maximizing condition $\partial I\left(\Gamma,t\right)/\partial t=0$ has no analytical solution, the ideal measurement time $t_{o}$ has to be found numerically and approximately determined by the more dominant decay among $\Gamma, \alpha$, see Fig.\ref{fig:fisher}(a). Alternatively, one can find the ideal measurement time heuristically from the characteristic time of the exponential decay by applying the condition $p\left(t\right)=1/e$, which marks the point at which the exponential decay becomes dominated by noise and which is the optimal measurement time for such an exponential decay \cite{Oviedo2020}. Doing so yields $t_{o}\approx0.5\alpha^{-2}\left(-\Gamma+\left(\Gamma^{2}+4\alpha^{2}\right)^{1/2}\right)$ which approximately scales as $min\left(\Gamma^{-1}, \alpha^{-1}\right)$. A common benchmark for precision is not the result achieved by a single measurement, but the best result achievable in an experiment of total duration time $T_{meas}\gg t_{o}$. Within this time period, the procedure is repeated $N=T_{meas}/\left(t_{o}+t_{m}\right)$ times, and we take the average result of all trials. Here, $t_{m}$ is the temporal overhead added to each trial and includes the duration of measurement and initialization \cite{sensing_review}. Using the additivity of the Fisher information and considering the Cramér-Rao bound \cite{Cramer1946, Rao1992} (which is saturated for the Bernoulli distribution), we get
\begin{equation} \label{precision_eq}
    \sigma_{\Gamma}=\sqrt{\frac{t_{o}+t_{m}}{T_{meas}}}\sqrt{\frac{1}{I\left(t_{o}\right)}}.
\end{equation}
In Fig.~\ref{fig:fisher} we show the SNR of the measurement, defined as $\Gamma/\sigma_{\Gamma}$, as a function of the decay rate, showing that an SNR greater than one is possible within the parameter range of a typical experiment. Precision and SNR can be further improved with more extensive dynamical decoupling, reducing the rate of the masking multiplicative decay.
Although these SNR values predict a feasible spectroscopy procedure, simply performing parametric modulation suffers from a significant loss of precision due to the additional decay. The above result is slightly lower compared to the SNR of simple exponential decoherence $p\left(t\right)=e^{-\tilde{\Gamma} t}$. Repeating the analysis above for the exponential rule yields the following
\begin{equation} \label{eq:fish_exp}
    I\left(\tilde{\Gamma},t\right)=\frac{t^{2}}{e^{\tilde{\Gamma}t}-1}\rightarrow t_{o}\approx\frac{1.6}{\tilde{\Gamma}}.
\end{equation}
Therefore the standard deviation of the estimator, with the same experiment and overhead times, is equal to
\begin{equation} 
    \sigma_{\tilde{\Gamma}}=\sqrt{\frac{1.5\tilde{\Gamma}}{T_{meas}}\left(\tilde{\Gamma}t_{m}+1.6\right)}\approx0.1\cdot \tilde{\Gamma},
\end{equation} 
giving an SNR value of around 10.

\section{\label{resolution}Resolution}
By analogy to the Rayleigh criterion in optics \cite{Rayleigh_crit}, we want to estimate the minimal distance that two spectral peaks must have in order to be distinguishable. First, we estimate the amount of information gained about a central frequency of a single-peak spectrum. From the spectroscopy procedure, we get an estimator for the spectrum at the modulation frequency, distributed normally by the central limit theorem: 
\begin{equation} \label{eq: decay_rate_prob}
    p\left(\hat{\Gamma}=x\right)=\frac{1}{\sigma_{\Gamma}\sqrt{2\pi}}\exp\left[-\frac{\left(x-\Gamma\right)^{2}}{2\sigma_{\Gamma}^{2}}\right],
\end{equation}
with $\sigma_{\Gamma}\propto \left(T_\phi T_{meas}\right)^{-1/2}$ calculated using the Cramér-Rao bound as specified in Appendix \ref{sec:SNR} and considering that the optimal measurement time for a single measurement is $t_o = T_\phi$, the FWHM of the power spectrum peak. The true value of the decay rate $\Gamma$ is equal to the spectrum of the noise sampled at the modulation frequency $\omega_{m}$. For the following analysis, we discuss a spectrum with a Gaussian shape around some frequency $\omega_{c}$; that is, 
\begin{equation}
    \Gamma=S\left(\omega_{m}\right)=\frac{A_{\omega}}{\sigma_{\omega}\sqrt{2\pi}}\exp\left[-\frac{\left(\omega_{m}-\omega_{c}\right)^{2}}{2\sigma_{\omega}^{2}}\right],
\end{equation}
where the uncertainty $\sigma_{\omega}\sim T_\phi^{-1}$ comes from the finite sharpness of our $sinc$ window function when we sample the spectrum, for more information see Appendix  \ref{appendix: decohernce-integrals} above. 
We would like to estimate the central frequency $\omega_{c}$. Substituting the spectral shape in Eq~(\ref{eq: decay_rate_prob}) yields the likelihood function
\begin{equation}
    p\left(\hat{\Gamma}=x\mid\omega_{c}\right)=\frac{1}{\sigma_{\Gamma}\sqrt{2\pi}}\exp\left\{-\frac{1}{2\sigma_{\Gamma}^{2}}\left[x-\frac{A_{\omega}}{\sigma_{\omega}\sqrt{2\pi}}e^{-\frac{\left(\omega_{m}-\omega_{c}\right)^{2}}{2\sigma_{\omega}^{2}}}\right]^{2}\right\}=\mathcal{L}\left(\omega_{c}\mid\hat{\Gamma}=x\right),
\end{equation}
Therefore, the Fisher information around $\omega_c$ is equal to 
\begin{equation}
\label{eq:FI w_c}
    I_{\omega_c}=\left\langle \left(\frac{\partial\log\mathcal{L}}{\partial\omega_{c}}\right)^{2}\right\rangle=\frac{A^2_{\omega}e^{-\frac{\left(\omega_{m}-\omega_{c}\right)^{2}}{\sigma_{\omega}^{2}}}}{2\pi\sigma_{\Gamma}^{2}\sigma_{\omega}^{6}}\left(\omega_{m}-\omega_{c}\right)^{2} \equiv \frac{A_{\omega}^{2}e^{-\frac{\delta\omega^{2}}{\sigma_{\omega}^{2}}}}{2\pi\sigma_{\Gamma}^{2}\sigma_{\omega}^{6}}\delta\omega^{2}\leq\frac{A_{\omega}^{2}}{2\pi e\sigma_{\Gamma}^{2}\sigma_{\omega}^{4}}.
\end{equation}
where we denote the detuning $\omega_{m}-\omega_{c}=\delta\omega$. The maximum information achievable from a single measurement [last inequality in Eq.~(\ref{eq:FI w_c})] is obtained when the modulation frequency is detuned one standard deviation from the central one, $\delta\omega=\pm\sigma_{\omega}$. A more realistic scenario is when the exact detuning is unknown, as a lower bound distribution, we assume that it is distributed uniformly within one $\sigma_{\omega}$ from the peak. Thus, the expected information contribution from a single modulation frequency experiment is 
\begin{equation} 
    \mathbb{E}\left(I_{\omega_c}\mid\delta\omega\sim U\left[-\sigma_{\omega},\sigma_{\omega}\right]\right)=\frac{A^2_{\omega}}{2\pi\sigma_{\Gamma}^{2}}\frac{e\sqrt{\pi}\Phi(1)-2}{4e\sigma_{\omega}^{4}}\simeq\frac{A^2_{\omega}}{10\pi\sigma_{\Gamma}^{2}\sigma_{\omega}^{4}}\approx\frac{1}{2}I_{\omega_c}^{\max},
\end{equation} 
where $\Phi\left(x\right)$ is the Gauss error function.

For a quantitative figure for the resolution of the method, we now turn to study the case of a two-peaked spectrum, for which
\begin{equation} \label{eq:likelihood_two}    \mathcal{L}\left(\omega^{\left(1\right)}_{c},\omega^{\left(2\right)}_{c}\mid\hat{\Gamma}=x\right)=\frac{1}{\sigma_{\Gamma}\sqrt{2\pi}}\exp\left\{ -\frac{1}{2\sigma_{\Gamma}^{2}}\left[x-\frac{A_{\omega}}{2\sigma_{\omega}\sqrt{2\pi}}\left(e^{-\frac{\left(\omega_{m}-\omega^{\left(1\right)}_{c}\right)^{2}}{2\sigma_{\omega}^{2}}}+e^{-\frac{\left(\omega_{m}-\omega^{\left(2\right)}_{c}\right)^{2}}{2\sigma_{\omega}^{2}}}\right)\right]^{2}\right\}.
\end{equation}

\begin{figure}
    \centering
\includegraphics[width=\textwidth,height=.4\textheight,keepaspectratio]{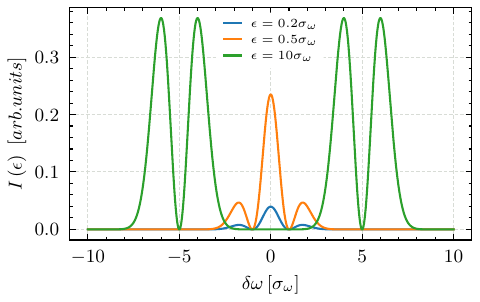}
    \caption{The different regimes of the Fisher information when estimating two peaks, as given by Eq.~(\ref{eq:FI_epsil}). In the resolvable regime where the distance of the peaks is larger than their width, $\epsilon\gg\sigma_\omega$, the Fisher information around each of the peaks is similar to the case of a single peak, Eq.~(\ref{eq:FI w_c}), and independent of the exact value of $\epsilon$. However, in the irresolvable regime $\epsilon\lesssim\sigma_\omega$, the maximal achievable information is at the mean frequency $\delta\omega=0$, and vanishes with $\epsilon\rightarrow0$, causing a divergence in the estimation precision.}
    \label{fig:resolution}
\end{figure}

Assuming the mean point of the two peaks $\bar{\omega}\equiv\left(\omega^{\left(1\right)}_{c}+\omega^{\left(2\right)}_{c}\right)/2$ is known, the parameter to be estimated is their spectral distance $\epsilon=\omega^{\left(2\right)}_{c}-\omega^{\left(1\right)}_{c}$. Rewriting Eq.~(\ref{eq:likelihood_two}) in these terms and evaluating the Fisher information around $\epsilon$ yields 
\begin{equation}
    \label{eq:FI_epsil}
    I_{\epsilon}=\left\langle \left(\frac{\partial\log\mathcal{L_\epsilon}}{\partial\epsilon}\right)^{2}\right\rangle =\frac{A_{\omega}^{2}}{8\pi\sigma_{\Gamma}^{2}\sigma_{\omega}^{6}}\left[e^{-\frac{\left(\delta\omega-\epsilon\right)^{2}}{2\sigma_{\omega}^{2}}}\left(\delta\omega-\epsilon\right)-e^{-\frac{\left(\delta\omega+\epsilon\right)^{2}}{2\sigma_{\omega}^{2}}}\left(\delta\omega+\epsilon\right)\right]^{2},
\end{equation}
where we denoted the detuning $\delta\omega=\omega_{m}-\bar{\omega}$. The Fisher information above has two different regimes, depending on the spectral difference $\epsilon$, as visualized in Fig.~\ref{fig:resolution}. In the \textit{resolvable} case where $\epsilon\gg\sigma_{\omega}$ the structure of Eq.~(\ref{eq:FI_epsil}) is of two separate peaks similar to the single frequency case, shifted by $\pm\epsilon$. Similarly, detuning of $\delta\omega=\pm\epsilon\pm\sigma_{\omega}$ maximizes the Fisher information.
\begin{equation}
    I_{\varepsilon}\left(\delta\omega=\pm\frac{\varepsilon}{2}\pm\sigma_{\omega}\right)=\frac{A_{\omega}^{2}}{2\pi\sigma_{\Gamma}^{2}\sigma_{\omega}^{4}}\left[e^{-\frac{1}{2}}-e^{-\frac{\left(1\pm\frac{\varepsilon}{\sigma_{\omega}}\right)^{2}}{2}}\left(1\pm\frac{\varepsilon}{\sigma_{\omega}}\right)\right]^{2}{\color{gray}\underset{\varepsilon\gg\sigma_{\omega}}{\underbrace{{\normalcolor \approx}}}}\frac{A_{\omega}^{2}}{2\pi e\sigma_{\Gamma}^{2}\sigma_{\omega}^{4}}.
\end{equation}
When the spectral distance is smaller, $\epsilon\lesssim\sigma_\omega$, the peaks become \textit{irresolvable}. The optimal detuning is 0 (meaning that the modulation frequency matches the mean point between the peaks) and the maximal achievable information is equal to 
\begin{equation} \label{eq:FI_epsilon}
    I_{\epsilon}\left(\delta\omega=0\right)=\frac{A_{\omega}^{2}}{8\pi\sigma_{\Gamma}^{2}\sigma_{\omega}^{6}}\left[e^{-\frac{\epsilon^{2}}{2\sigma_{\omega}^{2}}}\left(-\epsilon\right)-e^{-\frac{\epsilon^{2}}{2\sigma_{\omega}^{2}}}\left(\epsilon\right)\right]^{2}=\frac{A_{\omega}^{2}\epsilon^{2}}{2\pi\sigma_{\Gamma}^{2}\sigma_{\omega}^{6}}e^{-\frac{\epsilon^{2}}{\sigma_{\omega}^{2}}},
\end{equation}
leading to divergence in the spectral distance estimation  
\begin{equation}
    \Delta\left(\varepsilon\right)\simeq\left(\frac{2\pi}{A_{\omega}^{2}\epsilon^{2}T_{\phi}^{7}T_{meas}}\right)^{1/2}.
\end{equation}

\section{Leakage} \label{appendix: interaction_pic}
The quantum electrodynamics Hamiltonian of the tunable transmon circuit (without drive) is equal to
\begin{equation}
    H=4E_{c}n^{2}-E_{J,eff}\left(\varphi_{e}\right)\cos\left[\varphi-\varphi_{0}\left(\varphi_{e}\right)\right],
\end{equation}
where $n, \varphi$ are the charge (re-scaled to number of electrons) and SQ phase operators, $E_C=e^2/2C_\Sigma$ is the total capacitive energy of the circuit and $E_{J,eff}\left(\varphi_{e}\right)=\left(E_{J1}+E_{J2}\right)\sqrt{\cos^{2}\left(\varphi_{e}\right)+d^{2}\sin^{2}\left(\varphi_{e}\right)}$ is the effective and tunable Josephson energy. Its value depends on the asymmetry factor $d=\left(E_{J2}+E_{J1}\right) / \left(E_{J2}-E_{J1}\right)$, which also affects the phase shift $\varphi_{0}=\arctan\left[d\tan\left(\varphi_{e}\right)\right]$, and both are controlled by the external flux $\varphi_e$. The phase is canceled by the unitary shift $U_{\varphi0}=\exp\left(i\varphi_{0}n\right)$ leading to 
\begin{equation} \label{eq:transmon_rot_ham}
    \tilde{H}=4E_{c}n^{2}-E_{J_{eff}}\left(\varphi_{e}\right)\cos\left(\varphi\right)-\dot{\varphi}_{0}\left(\varphi_{e}\right)n,
\end{equation}
where the last term is non-negligible when we apply a fast flux drive. Around the potential well $\varphi=0$ the transmon is approximately an anharmonic oscillator with
\begin{equation}
    \varphi=\sqrt{\xi}\left(a^{\dagger}+a\right), \qquad
    n = \frac{i}{2\sqrt{\xi}}\left(a^{\dagger}-a\right),
\end{equation}
where $\xi=\sqrt{2E_C/E_J}$ is the small perturbation parameter from the exact QHO system. Hamiltonian Eq.~(\ref{eq:transmon_rot_ham}) in terms of ladder operators and $\omega_{T}\left(\varphi_{e}\right)\simeq\sqrt{8E_{J,eff}\left(\varphi_{e}\right)E_{c}}-E_{c}$ leads to the Hamiltonian in Eq.~(\ref{eq:full_cosine}) in the main text,
\begin{equation} \label{eq:Kerr_HamSM}
H=\omega_{T}\left[\varphi_{e}\left(t\right)\right]a^{\dagger}a-\frac{\eta}{2}a^{\dagger2}a^{2}-\frac{\dot{\varphi}_{0}\left[\varphi_{e}\left(t\right)\right]i}{2\sqrt{\xi}}\left(a^{\dagger}-a\right).
\end{equation}

\emph{Interaction picture} --- We show how the interaction Hamiltonian in Eq.~(\ref{eq:Kerr_Ham}) in the main text can be derived from Eq.~(\ref{eq:Kerr_HamSM}). 
We move to the interaction picture with respect to the Hamiltonian $H_{0}=a^{\dagger}a\cdot\left(\omega_{T}\left(\varphi_{e}\right)-\bar{\omega}_{T}\right)$. The unitary transformation does not affect the first and second terms as they commute with $H_{0}$, for the ladder operators $\exp\left(\beta a^{\dagger}a\right)a^{\dagger}\exp\left(-\beta a^{\dagger}a\right)=a^{\dagger}\exp\left(\beta\right)$ from the Baker-Hausdorff lemma. In our case, and within the quadratic approximation for the qubit's energy splitting,  
\begin{multline}
    \exp\left(\beta\right)=\exp\left\{i\intop_{t}\left[\omega_{T}\left(t_{1}\right)-\bar{\omega}_{T}\right]dt^{\prime}\right\}=\exp\left\{i\tilde{\omega}_{T}\intop_{t}\cos\left(2\omega_{m}t\right)dt^{\prime}\right\}=\\ \exp\left[i\frac{\tilde{\omega}_{T}}{2\omega_{m}}\sin\left(2\omega_{m}t\right)\right]=\sum_{n=-\infty}^{\infty}J_{n}\left(\frac{\tilde{\omega}_{T}}{2\omega_{m}}\right)e^{i2n\omega_{m}t},
\end{multline}
by the Jacobi Anger expansion. Furthermore, under the same approximation
\begin{equation}
    \dot{\varphi}_{0}\simeq d\dot{\varphi}_{e}=d\varphi_{ac}\omega_{m}\cos\left(\omega_{m}t\right)=\frac{d\varphi_{ac}\omega_{m}}{2}\left(e^{i\omega_{m}t}+e^{-i\omega_{m}t}\right).
\end{equation}
Substituting these results and organizing yields the desired result:
\begin{equation}
        H_{I}=\left\{ UHU^{\dagger}+i\dot{U}U^{\dagger}\right\} = \\ \bar{\omega}_{T}a^{\dagger}a-\frac{\eta}{2}a^{\dagger2}a^{2}-\frac{id\varphi_{ac}\omega_{m}}{2\sqrt{\xi}}\sum_{n}J_{n}\left(\frac{\tilde{\omega}_{T}}{2\omega_{m}}\right)\left[e^{i\left(2n+1\right)\omega_{m}t}a^{\dagger}-e^{-i\left(2n+1\right)\omega_{m}t}a\right].
\end{equation}

\section{\label{sec:simu_deets} Simulation details}
In this appendix, we elaborate on the considerations and procedures for the main results presented in this article, that is, spectrum reconstruction and leakage analysis. 

\subsubsection{Power spectral density reconstruction}
\begin{figure}[h]
\includegraphics[width=\textwidth,height=\textheight,keepaspectratio]{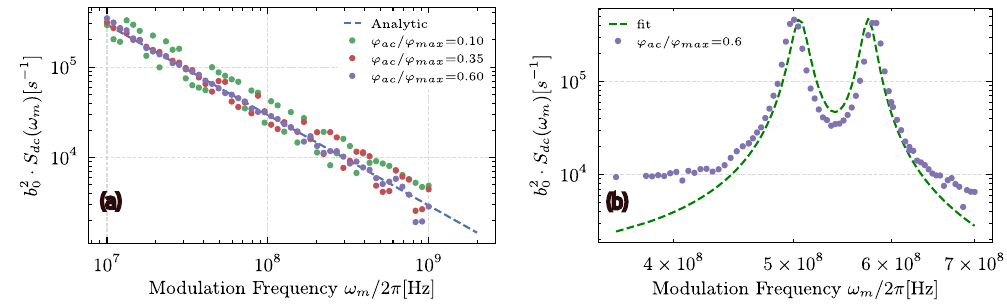}
      \caption{(a) Power spectral density reconstruction of 1/$f$ noise for a wide range of modulation frequencies, simulation parameters are equal to the main text simulation apart from generating only pink noise for the additive source. (b) Visualization of the fit that was performed to estimate the central points of the peaks. The data points are identical to Fig.~\ref{fig:PSD} in the main text, the dashed line shows the shape of the spectrum with the obtained parameters. }
    \label{fig:PSD_simu_details}
\end{figure}

In the reconstruction procedure, our first goal is to retrieve estimators for the power spectral density (PSD) at a certain frequency using the parametric spectroscopy method. Similarly to Fig.~\ref{fig:PSD} in the main text, in Fig.~\ref{fig:PSD_simu_details}(a) we present a reconstruction of the additive noise spectrum, in which the generated noise consisted of a simple 1/$f$ form, to show excellent performance on a wide frequency range. 

For the pink noise, we then add features to show the resolution of close frequencies in the high frequency region of the spectrum. They are created using an AR1 model \cite{Box1994}, in which each data point of the time series follows $x_{t}=\varphi x_{t-1} + \varepsilon_{t}$, where $\varepsilon_{t}$ is white noise. $\varphi$ is a parameter that controls the autocorrelation time of the sequence $T_{AR1}$ through $\varphi=\exp\left(-dt / T_{AR1}\right)$, where $dt$ is the minimal sample time spacing. The PSD of the process is equal to 
\begin{equation}
    S_{AR1}\left(\omega\right)=\frac{dt \sigma_{\varepsilon}^{2}}{1 + \varphi^2 - 2\varphi\cos\left(\omega dt\right)},
\end{equation}
approximating a Lorentzian peak with characterizing width of $T_{AR1}^{-1}$.
Fig.~\ref{fig:PSD_simu_details}(b) depicts the result of the fit made to obtain estimators for the central frequencies of the features. The noise model studied in the main text includes two shifted AR1 peaks and the pink shape. After obtaining spectral estimators at different frequencies, we fit the highest modulation amplitude samples (expected highest resolution by Eq.~(\ref{eq:FI_one_freq}) in the main text) to the spectral analytical shape. We estimate the values and confidence bounds of the free parameters $A_{\Phi}, \omega_{c}^{\left(i\right)}, T_{AR1}$ and the pink noise exponent, by mean value and std. from 100 successful fits ($R^2>0.8$). Since our focus is on the central frequencies of the peaks, we fit points within 200 MHz from them.

\subsubsection{Leakage Analysis}
For the leakage analysis results we simulate the Schrödinger equation using QuTIP solvers \cite{qutip}, under the Hamiltonian Eq.~(\ref{eq:transmon_rot_ham}) with a noise-free modulation signal $\varphi_{e}=\varphi_{ac}\sin\left(\omega_{m}t\right)$. We used 401 charge states, diagonalized the system numerically, and then projected onto the ten lowest eigenstates. We simulate modulation for $10\mu s$, and add eight Gaussian shaped pulses, equidistantly from one another during this period. The exact pulse shape incorporates the derivative removal by adiabatic gate (DRAG) and is given by
\begin{equation}
\left[\Omega_{x}\left(t\right)\sin\left(\omega_{d}t\right)+\Omega_{y}\left(t\right)\cos\left(\omega_{d}t\right)\right]n,
\end{equation}
with 
\begin{equation}
    \Omega_{x}\left(t\right)=Ae^{-\frac{\left(t-t_{c}\right)^{2}}{2\sigma_{p}^{2}}}, \qquad\Omega_{y}\left(t\right)=-\frac{\lambda}{\eta}\frac{\partial}{\partial t}\Omega_{x}\left(t\right).
\end{equation}
Anharmonicity $\eta/2\pi=300MHz$ and pulse width $\sigma_p=5ns$ with a cutoff of $4\sigma_p$ are used. The exact drive frequency, amplitude, and DRAG factor of the pulses $A/2\pi\approx34MHz,\omega_d/2\pi\approx6GHz,\lambda\approx0.47$, are obtained by minimizing average x gate infidelity at the sweet spot \cite{NIELSEN2002249}.
Finally, to estimate the leaked population at the end of the sequence, we take the maximal population out of the first two levels within the last 10 time steps of the simulation. The extended end period is taken to reduce sensitivity to phases of the active multi-level interactions during modulation.
\begin{figure}
    \centering
\includegraphics[width=\textwidth,height=.4\textheight,keepaspectratio]{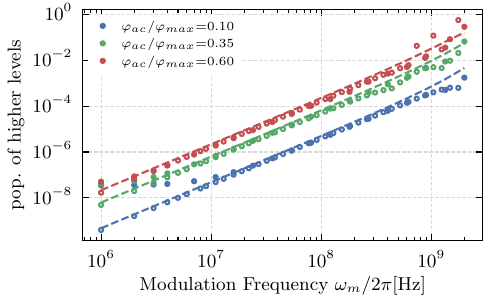}
    \caption{Kerr model accuracy. Results from simulations without dynamical decoupling pulses for transmon modulation using the full cosine (filled dots) and Kerr (empty dots) Hamiltonians. The dashed line represents the analytical model obtained in Eq.~(\ref{eq:leak_rabi}) in the main text. The main difference between the results is the \textit{plateau} of the full cosine potential at low frequencies, a result of varying $\varphi_e$ while simulating with operators obtained by numerical diagonalization at the sweet spot. The parameters of the transmon and analysis methods are the same as in Fig.~\ref{fig:leak_sim} in the main text.}
    \label{fig:leakage_SM}
\end{figure}


\section{\label{sec:correlation} Procedure to measure correlations in the environment}

This appendix describes the intriguing possibility of using parametric spectroscopy with two superconducting qubits to detect and measure spatial and temporal correlations in the environment. Here we follow the seminal idea published by von Lüpke et al. in Ref.~\cite{OliverCorrelations}. There, the authors use the spin-locking technique to detect correlations in the noise affecting two independent superconducting qubits. The coherent driving that controls the qubits is applied in the longitudinal ($\sigma_x$) axis, leading to the formation of a rotated basis in which transverse ($\sigma_z$) noise produces not only dephasing but relaxation as well. When the two qubits share the same environment, this leads to state mixing, from which the correlations in the noise can be inferred with adequate post-processing. Contrary to what spin-locking does, in the case of parametric spectroscopy, both the target noise and the driving occur in the transverse axis, so the same mixing that naturally happens with spin-locking is not immediately available. In this scenario, two possible solutions can be implemented in the case of parametric spectroscopy.

The first one has immediate possible implementation, and involves using the charge operator, rather than the flux operator, to drive the tunable transmons, as the charge operator acts on the longitudinal axis and would allow to mimic the scheme presented in \cite{OliverCorrelations}. In this case, no further theoretical development is needed, but there are two important things to bear in mind. On the one hand, the dynamical decoupling sequence that is applied concurrently with the parametric modulation, now will need to be applied through the flux channel, contrary to conventional parametric spectroscopy, where the driving goes in the flux channel and the DD sequence in the charge channel. This means, additionally, that the dynamical decoupling pulses address the rotated basis. Additionally, driving the transmons through the charge channel comes at the price of increasing the noise affecting the qubits, which would reduce the coherence time gain which is a major advantage of parametric spectroscopy. Moreover, such an experiment would encounter similar problems to what spin-locking finds when tackling the high frequency end of the spectrum. Namely, due to the strong Rabi drivings required, the system becomes more prone to leakage. Both effects can still be mitigated with the dynamical decoupling sequences that are applied concurrently with the driving, but the efficiency of the spectrometer will decrease with respect to what we find in the present article for single qubit spectrometry.

The alternative route that we envision to induce mixing of the qubits and therefore sensitivity to correlations would be to longitudinally couple the tunable transmons through, e.g., a capacitor, which would play the role of rotating the bases of the qubits. In this case, further theoretical development is required, as the rotated basis which is now used to measure correlations is time dependent, and leads to similar expressions as those analyzed when studying the leakage. Meaning that, although no experimental limitation exists in principle, a full understanding of the behavior of the energy levels and the mixing in terms of Bessel functions is required, which leads to the necessary design of appropriate dynamical decoupling sequences via, e.g., optimization of the pulses.

We would like to note that in both of the cases described above, either using charge driving or coupling the qubits, the advantage that parametric spectroscopy has is twofold: The use of tailored dynamical decoupling sequences which increase the coherence time and reduce the leakage, leading to improved resolution, and the flexibility to choose the rotation angle, or degree of mixing, which can be used to optimize the sensitivity to correlations balanced with controlling the degree to which noise and leakage are augmented.

\section{\label{sec:t1_effect} Relaxation effects}
\begin{figure}[h]
\includegraphics[width=\textwidth,height=\textheight,keepaspectratio]{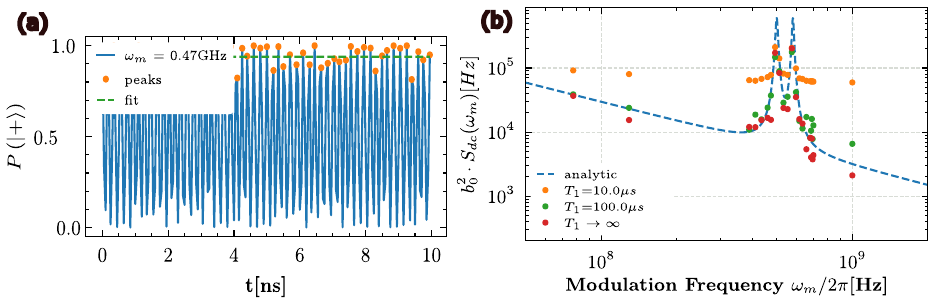}
         \caption{Visualization of different $T_1-T_{\phi}$ ratios and the resulting peak separation performance. (a) Noise estimating procedure. The blue signal is the average of $2^8$ traces of $\left|+\right>$ state probability during evolution (first $10ns$) governed by the master equation Eq.~(\ref{eq: t1 master}),  with parametric modulation parameters $\varphi_{ac}=0.6\cdot\varphi_{max}; \omega_m \approx 470MHz$ and stochastic flux noise. The orange dots show the extracted peaks and the result of the exponential fit in dashed green. (b) spectrum reconstruction with different relaxation rates. The dots represent exponential decay rates extracted as described in (a), with high modulation amplitude $\varphi_{ac}=0.6\cdot\varphi_{max}$, and different frequencies $\omega_m$. The different colors (orange, green, and red) represent different $T_1$ values for the simulations $\left(10\mu s,100\mu s,\infty\right)$ respectively.} \label{fig:T1}
\end{figure}

In this appendix, we discuss the effects of a finite $T_1$ relaxation time on parametric spectroscopy. As relaxation and dephasing are separate noise processes affecting the qubit, relaxation is hindering the protocol in the case where $T_1\lesssim
 T_\phi$, in which phase information is lost due to decay to the ground state before dephasing can be analyzed. This limitation can be pushed to lower values if we post-select only the sample that did not relax, for a reasonable retaining ratio. In systems with $T_1\ll T_{\phi}$ parametric spectroscopy, like any other dephasing measurement method, cannot be utilized. In Fig.~\ref{fig:T1}, we present the results of parametric spectroscopy of a qubit subject to different rates of relaxation. To extract these dephasing rates we simulated the qubit (2-level) master equation \cite{qutip}
 \begin{equation}
    \label{eq: t1 master}
    \dot\rho{\left(t\right) = -i\left[\rho\left(t\right), H\left(t\right) \right]} + \frac{1}{2T_1}\left[2\sigma^-\rho\left(t\right)\sigma^+ - \rho\left(t\right)\sigma^+ \sigma^- - \sigma^+\sigma^- \rho\left(t\right)\right],
\end{equation}
with the parametric spectroscopy Hamiltonian
\begin{equation}
H=\omega_{T}\left[\varphi_e\left(t\right)\right]\frac{\sigma_{z}}{2},
\end{equation}
under modulation $\varphi_e\left(t\right)$ as in Eq.~(\ref{eq:flux_drive}) in the main text and a Lindbladian term for $T_1$ relaxation. Stochastic additive noise with a spectrum identical to Fig.~\ref{fig:PSD} in the main text is added to the flux signal, whereas multiplicative noise and DD pulses are not included in these simulations for simplicity. The system is initialized in the $\left|+\right>$ state, and we measure its projection probability during evolution. The peaks of the oscillatory signal are extracted and fitted to an exponential decay signal, yielding decay rates for the spectrum reconstruction. It is observed that $1/T_1$ manifests a lower bound for the smallest spectral density that can be measured using the method. For values of $T_1$ larger than the typical measured dephasing rates, this bound is low enough, and parametric spectroscopy provides a spectral peak resolution similar to the $T_1\rightarrow\infty$ case in the main text. However, for higher relaxation rates, the extracted spectral densities cling to this lower bound and harm the procedure. This degrading effect can be mitigated, by post-selecting the non-relaxed samples only, with a price of extending $T_{meas}$ and reducing resolution.

\end{document}